\documentstyle[aasms4,psfig]{article}
\received{April 2, 1999}
\accepted{June 11, 1999}
\lefthead{Thi et al.}
\righthead{H$_2$ emission from the GG Tau binary system}
\journalid{}{}
\articleid{}{}

\def\grapprox{$_>\atop{^\sim}$}

\begin{document}
\title{Detection  of H$_{2}$  pure rotational  line emission from  the  GG~Tau 
binary system\footnotemark}
\footnotetext{Based on observations with {\it ISO}, an ESA project 
with instruments funded by ESA Member States (especially the PI countries : 
France, Germany, the Netherlands, and the United Kingdom) and 
with participation of ISAS and NASA.}

\author{Wing-Fai Thi\altaffilmark{2},  Ewine F.\ van Dishoeck\altaffilmark{2},
Geoffrey  A.\  Blake\altaffilmark{3}, Gerd-Jan  van  Zadelhoff\altaffilmark{2},
Michiel R.\  Hogerheijde\altaffilmark{4}}
\altaffiltext{2}{Leiden Observatory, P.O. Box 9513, 2300 Leiden, 
The Netherlands.} 
\altaffiltext{3}{Division of Geological \& Planetary Sciences, 
California Institute of Technology 150--21,   Pasadena,   CA  91125,   USA.}
\altaffiltext{4}{Department  of   Astronomy,  601   Campbell  Hall,
University of California, Berkeley, CA 94720-3411, USA.}

%\slugcomment{TEXT}

%%%%%%%%%%%%%%%%%%%%%%%%%%%%%%%%%%%%%%%%%%%%%%%%%%%%%%%%%%%%%%%%%%%%%%%%%%%%%% 
% abstract

\begin{abstract}
We  present  the first  detection  of  the  low-lying pure  rotational
emission lines of H$_2$ from circumstellar disks around T~Tauri stars,
using  the Short Wavelength  Spectrometer on  the {\it  Infrared Space
Observatory}. These lines provide a direct measure of the total amount
of warm molecular  gas in disks.  The $J$=2$\to$0  S(0) line at 28.218
$\mu$m  and the  $J$=3$\to$1  S(1)  line at  17.035  $\mu$m have  been
observed toward the double binary system GG~Tau.  Together with limits
on the $J$=$5\to3$ S(3) and $J=7\to5$ S(5) lines, the data suggest the
presence of gas at $T_{\rm  kin}\approx 110\pm 10$  K with a mass of
$(3.6\pm 2.0)\times  10^{-3}$  M$_\odot$ ($\pm 3 \sigma$). This amounts
to $\sim 3$~\% of the total gas  + dust mass of  the circumbinary disk
as imaged by millimeter interferometry, but is larger than the estimated
mass of the circumstellar disk(s).  Possible origins for the warm 
gas seen in H$_2$ are discussed in terms of photon and wind-shock
heating mechanisms of the circumbinary material, and comparisons with 
model calculations are made.

\end{abstract}

% ----------------------------------------------------------------
% keywords

{}

%%%%%%%%%%%%%%%%%%%%%%%%%%%%%%%%%%%%%%%%%%%%%%%%%%%%%%%%%%%%%%%%%%%%%%%%%%%%%%%
\section {INTRODUCTION}
T~Tauri stars  are considered to resemble our  Sun at an age  of a few
million years. Studies of their surrounding gas and dust can therefore
provide important  clues on the  early evolution of the  solar nebula.
It  is well  established through  surveys at  infrared  and millimeter
wavelengths  that most  T~Tauri  stars have  circumstellar disks  with
masses  of $\sim  10^{-3} -  10^{-1}$ M$_{\odot}$  and sizes  of $\sim
100-400$~AU (see overviews by \cite{BS96}, Dutrey et al.\ 1996, Mundy et
al.\ 2000). In  addition to serving as a conduit for mass accretion onto
the young star, the disks also  provide a reservoir of gas and dust for
the  formation  of potential  planetary  systems  (Shu  et al.\  1993).
Theories of disk evolution depend  strongly on the radial and vertical
temperature structure of the disks  (e.g., Hartmann et al.\ 1998), but
these parameters are still poorly constrained by the available observations.

We report  here the results of a  deep survey for the  lowest two pure
rotational lines of H$_2$, the  $J$=$2\to0$ S(0) line at 28.218 $\mu$m
and the $J=3\to1$ S(1) transition at 17.035 $\mu$m, using the Short Wavelength
Spectrometer  (SWS)  on board  the  {\it  Infrared Space  Observatory}
(ISO). In emission, the H$_2$ lines originate from levels at 509.9~K and 
1015.1~K above ground and are thus excellent tracers of the ``warm'' 
($T${\grapprox}80~K) gas in disks, especially in the interesting inner 
part where Jovian planets may form.  H$_2$ has the advantage that it
dominates the mass budget and that it does not deplete onto grains,
contrary to  CO.  Moreover,  the lines are  optically thin up  to very
high  column   densities  owing  to   the  small   Einstein  $A$
coefficients for electric quadrupole transitions, so that the modeling
of the  radiative transfer  is simple.  

Here we present H$_2$ pure rotational line observations of the GG Tau
system, which is situated at the edge of the Taurus-Auriga  cloud complex
at a distance of approximately 140 pc (\cite{KDH}).  GG~Tau consists of
two close binary pairs separated by 10$''$, or 1400 AU. The main binary 
GG~Tau A has a separation of $\sim 35$~AU (Ghez  et al.\ 1997)  and is  
composed of a  K7 and a  M0.5 star (White   et  al.\   1999),   both  
classified as emission-line  or ``classical'' T~Tauri stars by Herbig \& 
Bell (1988) with an estimated accretion rate of 2 $\times$ 10$^{-8}$ 
M$_{\odot}$ yr$^{-1}$ (Hartmann et al.\ 1998).  The GG  Tau B binary 
is comprised of an M5 and M7 star separated by $\sim  200$ AU. The age
of the GG Tau system is estimated to be $\sim 1.5$ Myrs by White et 
al.\ (1999), using the evolutionary  models of Baraffe et al.\  (1998) 
and assuming the four stars to be  coeval.  

High spatial resolution images taken in the near infrared show  that each of
the stars in the GG  Tau A system has associated circumstellar  material
within a radius of  less than 10~AU (Roddier et  al.\ 1996).  These stars are 
located within a  cavity of radius $\sim 200$ AU cleared by the dynamical
interaction of the binary (Ghez et  al.\ 1997), and a  circumbinary disk
extending  up to $\sim 800$  AU as imaged  at  millimeter  wavelengths 
(Dutrey  et  al.\  1994, Guilloteau  et  al.\ 1999).  The 
circumbinary  disk  mass of $\sim 0.12$  M$_{\odot}$,  
deduced  from  the  strong 
millimeter  dust continuum  and assuming  a  gas-to-dust ratio  of  100:1, is 
one of  the largest observed  to date (Guilloteau  et al.\ 1999). 
However, observations of $^{13}$CO and C$^{18}$O indicate gas masses that are
up to a factor of 100 lower,  (Dutrey  et al.\  1994).  Explanations for this
discrepancy include  the possible freeze-out  of CO in the  cold outer part of
the  disk and/or a gas dissipation time  scale that is shorter than  that  of 
 the  dust  (Zuckerman  et  al.\   1995).   The  H$_2$ observations presented 
here allow a direct measurement  of the amount of warm gas in the disk.

The H$_2$ data  for GG~Tau  form part of  a survey of  a larger number  of T
Tauri and Herbig Ae stars with  the ISO-SWS by Thi et al.\ (1999a; and
in preparation).  An initial account  of the results has been given in
van Dishoeck et al.\ (1998).

%%%%%%%%%%%%%%%%%%%%%%%%%%%%%%%%%%%%%%%%%%%%%%%%%%%%%%%%%%%%%%%%%%%%%%%%%%%%%
\section{OBSERVATIONS AND DATA REDUCTION}

The low-lying  pure-rotational H$_{2}$  $J=2\to0$ S(0) line  at 28.218
$\mu$m and the $J=3\to1$ S(1) line at 17.035 $\mu$m were observed with
the ISO-SWS  in the AOT02  mode (\cite{Gra}).  The observations  were 
centered in the direction of GG~Tau  A at RA(2000)=04$^h$32$^m$30$^s$, 
DEC(2000)=17$^{\circ}$31$'$42$''$.  Typical
integration times were 600--1000~s per line, in which the 12 detectors
were  scanned several  times  over the  28.05--28.40 and  16.96--17.11
$\mu$m ranges  around the lines. The  $J=5\to 3$ S(3)  9.66 $\mu$m and
$J=7\to 5$ S(5)  6.91 $\mu$m lines were measured  in parallel with the
S(0) and  S(1) lines, respectively,  at virtually no extra  time.  The
spectral resolution for point sources  is 2000 (150 km s$^{-1}$) at 28
$\mu$m, 2400 (125 km s$^{-1}$) at 17 $\mu$m, 2280 (130 km s$^{-1}$) at
9.7  $\mu$m  and  1550 (195  km  s$^{-1}$)  at  6.9 $\mu$m.   The  SWS
apertures are $20''\times  27''$ at S(0), $14'' \times  27''$ at S(1),
and $14''\times 20''$ at the S(3) and S(5) lines. Two independent sets
of observations have  been carried out in orbital  revolutions 668 and
834.  The expected  peak fluxes of  the H$_2$ lines  are close to
the sensitivity limit of the instrument,  and the raw data show a high
level  of noise induced by charged-particle impacts  on the
detectors.  In order  to  extract the  H$_2$  lines, special  software
designed to  handle weak  signals was used  for the data  reduction in
combination  with  the  standard  Interactive  Analysis  Package.  The
details  and justification  of the  methods used  in the  software are
described  by  Valentijn \&  Thi  (1999).   Because the mid-infrared continuum
emission from GG Tau is weak, $<$1 Jy at $<$17 $\mu$m and $\sim 3$ Jy
at 28 $\mu$m, the data do not suffer from fringing effects caused by an
inadequate responsivity function correction.

%%%%%%%%%%%%%%%%%%%%%%%%%%%%%%%%%%%%%%%%%%%%%%%%%%%%%%%%%%%%%%%%%%%%%%%%%%
\section{RESULTS}

The H$_2$ S(0) and S(1) lines are detected in both sets of observations 
(see Table~1) and the differences  in fluxes between the two sets are 
$\sim 10$\%. This is well within the estimated total error of $\sim 30$\%, 
which is mostly due to uncertainties in the flux calibration. The S(3) 
and S(5) lines are not detected down to a limit of 
$\sim (5-7)\times 10^{-15}$ erg s$^{-1}$ cm$^{-2}$ ($3\sigma$). 
The  spectra of  the  H$_{2}$ S(0)  and  S(1) lines  are displayed  in
Figure~1  for  revolution  668.  Since the  turbulent  and  rotational
velocities in disks are only a few km s$^{-1}$, respectively, both lines
are unresolved. After subtraction of the continuum, the lines are fitted
by Gaussians with  widths fixed by the instrumental resolution
and the line fluxes  are computed from $-3 \times$ to 
$+3 \times$ HWHM. 

In the  optically thin  limit, the observed  line fluxes  are directly related
 to the  populations in  the H$_2$  $J$=2 and  3  levels.  The derived kinetic
temperature assuming local thermodynamic equilibrium (LTE)  is $110 \pm 10$~K,
where the error bar reflects the 30\% uncertainty in the fluxes.  Although the
$J$=3  level has a factor  of 40  lower population  than the $J$=2 level  in
gas  with a temperature near 100~K, the  radiative transition  probability of
the S(1)  line, $A_{31}=4.8\times  10^{-10}$ s$^{-1}$, is  much larger than 
that  of   the  S(0)  transition, $A_{20}=2.9\times  10^{-11}$ s$^{-1}$.   In 
addition,  the  spectral resolution  at  17~$\mu$m  is somewhat higher than
that at 28  $\mu$m, and the line-to-continuum ratio is correspondingly larger.
Both of these factors explain why the S(1) line is detectable as well. The
limits  on the  S(3) and S(5)  lines imply temperatures  less than 
$\sim 260$~K
and $\sim 450$~K, respectively, neglecting any correction for differential
extinction.

Since the  ISO beam is much  larger than the size  of the circumbinary
disk,  it is  important to  check whether  any of  the  observed H$_2$
emission may arise from  residual extended envelope or cloud material.
H$_2$ lines up to S(9)  are readily detected toward embedded Herbig Ae
stars with the  ISO-SWS (e.g., van den Ancker et  al.\ 1998). In these
cases, the emission is dominated  by the interaction of the young star
with its surrounding envelope  through shocks and ultraviolet photons.
The  typical  H$_2$  excitation  temperatures for  these  regions  are
$T_{\rm exc}$=500--700~K,  much larger than the value found for GG Tau.
Deep ISO searches for the H$_2$ S(0) and S(1) lines  toward
diffuse  and translucent  clouds have  been performed  by Thi  et al.\
(1999b).  The lines are not detected down to $2\times 10^{-14}$ erg s$^{-1}$
cm$^{-2}$ ($2\sigma$) in gas with densities of a few hundred to a few thousand
cm$^{-3}$ exposed to the normal  interstellar radiation  field. The $^{12}$CO 
1--0 emission around GG~Tau  is less than 50 mK  (3$\sigma$) (\cite{Skr}),
more than an order of magnitude lower than found for diffuse clouds such as
that toward  $\zeta$~Oph. The corresponding mass in the SWS beam is
estimated to be less than a few $\times 10^{-4}$ M$_{\odot}$, significantly
lower than the mass derived from the H$_2$ lines (see below).

Keck images of GG Tau in $K'$ continuum and H$_2$ 
emission in the $v$=1$\to$0 and $v$=2$\to$1 S(1) lines at 2.1250 and 
2.2486 $\mu$m were obtained on 03 and 06 November 1998 using the 
facility Near InfraRed Camera (NIRC) and the appropriate filters. No H$_2$
emission down to $\sim 20$ $\mu$Jy (2$\sigma$) was detected in a 1.5-6$''$
(i.e. 200-800 AU) radius around the stars, nor
outside this region.  This translates to a limit
on the intensity of $\sim 3\times 10^{-6}$ erg s$^{-1}$ cm$^{-2}$ sr$^{-1}$.
Altogether, we are confident that the bulk of the H$_2$
emission toward GG Tau originates from the disk(s) rather than
interstellar material  in  the ISO  beam.   

Assuming  no continuum extinction and LTE  excitation, the corresponding mass
of  warm gas is computed via the relation
$$
M_{\rm warm\ gas}=  1.76\times10^{-20} \ {{F_{ul}\ d^2} \over
{(h\nu_{ul}/{4\pi})}\ A_{ul}\ x_J(T)}\ \ {\rm M_{\odot}}~~~~~,
$$
where $F_{ul}$ is the integrated flux in erg s$^{-1}$ cm$^{-2}$, $d$
is the distance of GG~Tau in pc (taken to be 140 pc), $\nu_{ul}$ is
the frequency of the transition in Hz, and $x_J$($T$) is the
fractional population in the upper rotational level $J_u$. The derived
amount of warm gas is (3.6 $\pm$2.0) $\times$ 10$^{-3}$ M$_{\odot}$
(3$\sigma$), including the 30\% uncertainty in the fluxes.  The
derivation assumes that the ortho/para H$_2$ ratio is $\sim 1.8$, the
LTE value at 110~K.  If the emission were affected by 30 magnitudes of visual
extinction, the derived excitation temperature would increase to 121 K
and the mass to $4.0 \times$ 10$^{-3}$ M$_{\odot}$.

%%%%%%%%%%%%%%%%%%%%%%%%%%%%%%%%%%%%%%%%%%%%%%%%%%%%%%%%%%%%%
\section{DISCUSSION}

The inferred warm H$_2$ mass of $\sim 3.6\times$ 10$^{-3}$ M$_{\odot}$ is
about 3\% of the total gas + dust mass of 0.12 M$_{\odot}$ derived from
millimeter continuum observations of the circumbinary disk assuming a
gas-to-dust ratio of 100:1 and a gas + dust absorption coefficient of 0.01
cm$^2$ g$^{-1}$ at 2.6 mm (Guilloteau et al.\ 1999).  The temperature is much
higher than that derived from the continuum spectral energy distribution and
optically thick CO emission, which give a temperature of only 34~K at the
inner edge of the circumbinary disk at 180 AU.

Where does the warm H$_2$ emission originate and what is the heating
mechanism?  The bulk of the circumbinary disk is too cold to account for
emission by gas at 100 K.  Moreover, the disk is optically thick in the
mid-infrared continuum. The H$_2$ emission must therefore arise either from
the circumstellar disk(s), or from the surface  layers and inner edges of the
circumbinary disk.  Two kinds of heating mechanisms may be at work: heating by
absorption of part of the stellar and accretion luminosity, and heating by
dynamical processes including shocks and turbulent decay. These possibilities
are discussed in turn below.

Several radiative processes must be examined. The first possibility is
that the H$_{2}$ lines arise from material within $10-20$~AU around
the individual stars, where the gas and dust are heated by the
ultraviolet radiation from the YSOs.  In general, a large fraction of
this warm gas in the inner circumstellar disk(s) may be hidden by the
optically thick continuum of colder surrounding dust, especially if
the disks are observed nearly edge on. However, GG Tau presents a
special case, since the dynamical interaction of the binary has
cleared the inner part of the circumbinary disk.  The near-infrared
observations of Roddier et al.\ (1996) suggest that at least some of
the radiation from the inner disks can escape through holes in the
circumbinary disk combined with favorable orientations of the
material.  The masses of the inner circumstellar disks are estimated
to be only $\sim 10^{-4}$ M$_{\odot}$ each, however, based on the
millimeter continuum data (Guilloteau et al.\ 1999). It therefore does
not appear that there is sufficient mass in the inner disks to explain
the warm H$_2$ emission. 

Consider next the case of the more extended circumbinary disk. If this
disk is flared, as is expected if hydrostatic equilibrium is approached,
there exists a surface layer that is heated by
radiation from the central star(s) to temperatures near 100 K out to
radii of $\sim 100$ AU (e.g., Chiang \& Goldreich 1997, 1999). These disk
models have a non-isothermal vertical temperature profile, and the
warm gas is located in the near-surface regions of the disk. Thus, the
emission arising from the heated material is not absorbed by cooler
layers before reaching the Earth. An H$_2$ excitation calculation has
been performed using the density and temperature structure of the
standard model of Chiang \& Goldreich (1997), assuming a gas-to-dust
mass ratio of 100:1 with $T_{\rm gas}=T_{\rm dust}$.  Typical S(0) and
S(1) fluxes from a single face-on surface layer are $7\times
10^{-16}$ and $2 \times 10^{-15}$ erg
s$^{-1}$ cm$^{-2}$, respectively. These values depend sensitively on the
adopted continuum opacities at mid-infrared wavelengths and the geometry,
resulting in uncertainties of factors of 2--3. Even taking these factors into
account, however, the model fluxes are a factor of 5--10  lower than the
observed values. Moreover, the model S(0)/S(1) ratio of 0.4 is lower than the
observed ratio of $1.0\pm0.5$, and the model S(3)/S(1) ratio of 0.4 is higher
than  the observed ratio of $<$0.2. A more detailed radiative transfer
simulation including the disk inclination angle of $\sim
35^{\circ}-43^{\circ}$ (Roddier et al.\ 1996, Guilloteau et al.\ 1999) and
possible velocity shifts between the warm and cold gas is needed to provide a
more accurate assessment of this model, but is beyond the scope of this Letter.

A third radiative mechanism may be provided by ultraviolet radiation from the
star-inner disk boundary layer(s) which can irradiate the inner edge of the
circumbinary disk (or any residual gas in the cavity) and heat it to $\sim
100-200$~K. This situation has been described for  circumstellar envelopes by
Spaans et al.\ (1995). The intensity of the ultraviolet radiation from a
10,000~K boundary layer with a luminosity of 0.3$L_{\rm bol}$ is estimated to
be up to a factor of 600 larger than the average interstellar radiation field
at a distance of $\sim 180$ AU. If the density at the boundary is assumed to
be a few$\times 10^6$ cm$^{-3}$, the computed H$_2$ fluxes are $\sim 3\times
10^{-15}$ and $\sim 2\times 10^{-15}$ erg s$^{-1}$ cm$^{-2}$ for the S(0) and
S(1) lines, respectively, an order of magnitude lower than observed. Similar
discrepancies between models and H$_2$ observations are found for dense
interstellar clouds exposed to ultraviolet radiation (e.g., Draine \& Bertoldi
1999), indicating that the heating mechanisms are not fully understood. The
S(0)/S(1) ratio of $\sim 1.5$ and S(3)/S(1) ratio of $< 10^{-3}$ in these
models are consistent with the data within the errors, however. Further
modeling of the radiative heating of the surface layer and inner edge of the
circumbinary disk is needed to investigate whether a combination of
these radiative mechanisms can reproduce the observations.

An alternative class of heating mechanisms involves shocks caused by infalling
material at the inner disk surface(s) or by the interaction between an
outflowing supersonic wind and the surfaces of the circumstellar or
circumbinary disk(s) (Hartmann \& Raymond 1989). Evidence for such winds comes
from optical observations of atomic and ionic lines, from which Hartigan et
al.\ (1995) derive a mass loss rate of 7.9 $\times$ 10$^{-10}$ M$_{\odot}$
yr$^{-1}$ for GG Tau. Since the masses of the inner circumstellar disk(s) are
too low to explain the H$_2$ emission, only the interaction of the wind with
the circumbinary disk at $\sim 180$~AU needs to be considered. The problem
with these models is that shocks are expected to warm the surface layers to
sufficiently high temperatures to emit strongly in the S(5) and S(3) lines and
the 2 $\mu$m vibration-rotation lines.  Consider as an example the wind-disk
models of Hartmann \& Raymond (1989). For typical wind velocities of 200 km
s$^{-1}$, the estimated shock velocities along the disk surface range from
20--40 km s$^{-1}$ at distances of 50--200 AU.  Comparison with the $J$-- and
$C$--shock models of Burton, Hollenbach \& Tielens (1992) and Kaufman \&
Neufeld (1996) shows that in virtually all models the flux in the S(3) line is
predicted to be larger than that in the S(1) line, in contrast with the
observations. Using the S(3)/S(1) ratio as a constraint, at most 30\% of the
S(1) emission could be contributed by shocks. The lack of detected H$_2$
$v$=1$\to$0 S(1) emission also indicates the absence of shocks faster than
$\sim 20$ km s$^{-1}$.  Thus, the H$_2$ S(3) and 2 $\mu$m upper limits suggest
that heating by shocks is unlikely to be the major contributor to the line
emission.  Most likely, a combination of heating by ultraviolet photons and
dynamical processes is responsible for the warm molecular gas.

In summary,  this work demonstrates  that H$_2$ pure  rotational lines
can be  detected from  disks around pre-main  sequence stars  and that
they provide complementary  information to sub-millimeter observations
of  CO and other  molecules. The  H$_2$ observations  are particularly
sensitive to the warm gas in the disks. With current instrumentation, 
masses of warm H$_2$ can be detected that are only a small fraction of the
total gas + dust mass in circumstellar disks. The line ratios
provide important constraints on the heating mechanisms. In order for the H$_2$
emission to escape, however, the emission must arise either from the disk
surface layers  or requires the presence of gaps or holes in the
disks. In the case  of GG~Tau,
the binary nature  of this system has cleared a large inner
cavity in the circumbinary disk, which may have facilitated the
detection of  the lines.  Future  observations at higher spectral and
spatial resolution  such as provided by  mid-infrared spectrometers on
ground-based telescopes and aboard platforms such as the {\it Stratospheric
Observatory For Infrared Astronomy} (SOFIA) and the {\it Next Generation
Space Telescope} (NGST), 
accompanied  by more sophisticated modeling, should be
able to clarify  the origin of the H$_2$ emission  from disks around T
Tauri and Herbig Ae stars and allow much more sensitive searches.  

%%%%%%%%%%%%%%%%%%%%%%%%%%%%%%%%%%%%%%%%%%%%%%%%%%%%%%%%%%%%%%%%%%%%%%%
\acknowledgments{This   work   was   supported  by   the   Netherlands
Organization for  Scientific Research (NWO)  grant 614.41.003, and
by grants to GAB from NASA (NAGW-4383 and NAG5-3733).  MRH is supported
by the Miller Institute for Basic Research in Science.  The W.M.~Keck
Observatory is operated as a  scientific partnership between Caltech, 
Univ.~of California, and NASA.  It was made possible by the generous 
financial support of the W.~M.~Keck Foundation.}

\clearpage

\figcaption[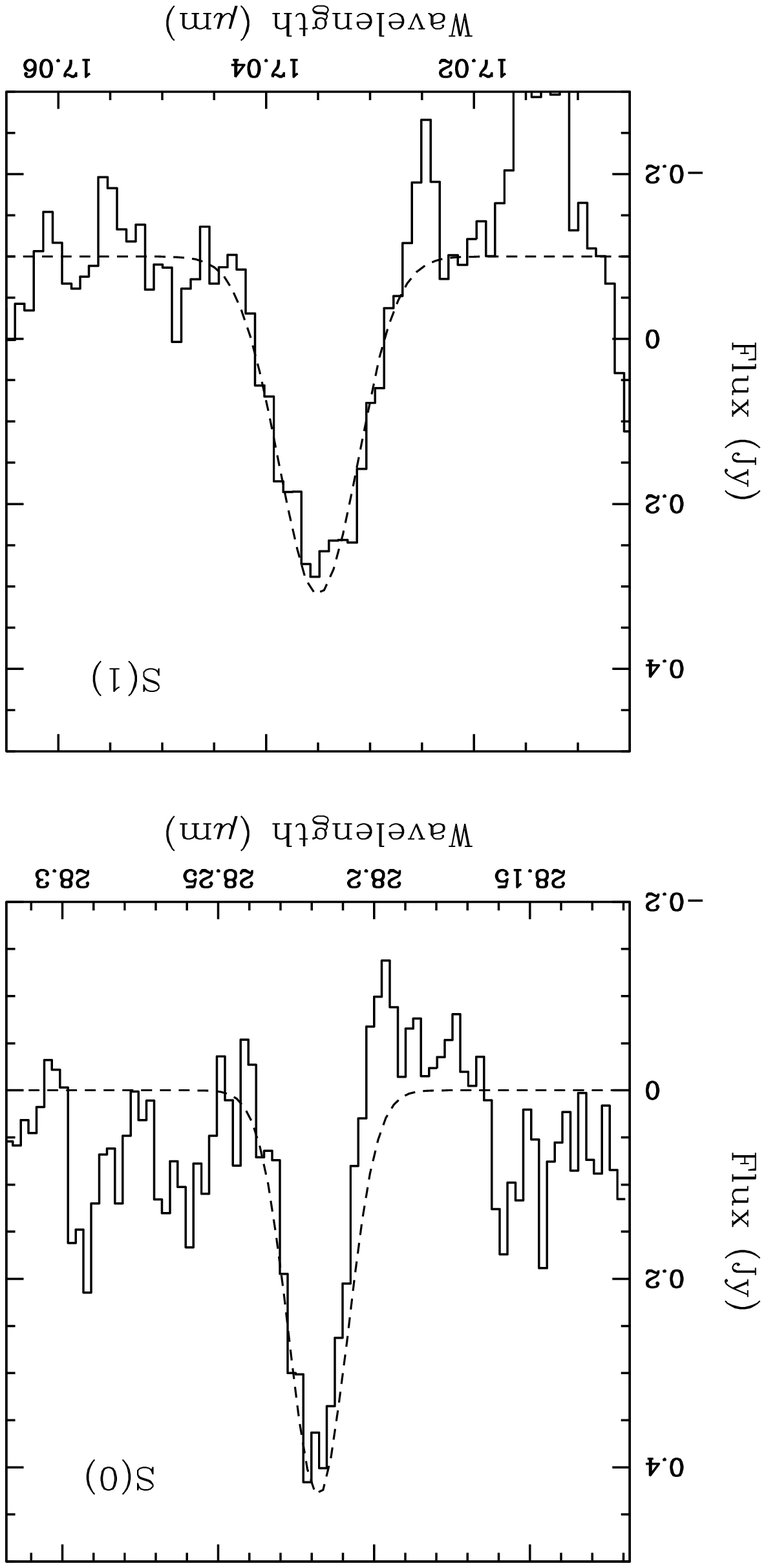]{H$_2$ $J=2\to 0$ S(0) (top panel) and S(1) 
$J=3\to 1$ (bottom panel) emission 
toward GG~Tau obtained with the ISO-SWS after subtraction of the continuum. 
The solid lines indicate Gaussian fits to the data with a width
fixed at the instrumental resolution.}

% --------------------------------------------------------------------------
\begin{deluxetable}{cccc}
\tablecaption{}
\tablewidth{0pt}
\tablehead{
\colhead{Revolution} & \colhead{Transition} & \colhead{Rest wavelength} 
&\colhead{Integrated Flux}\nl
                     &                      &  \colhead{$\rm \mu$m} 
& \colhead{erg s$^{-1}$ cm$^{-2}$}}
\startdata
668 & H$_2$ 0-0 S(0) & 28.218  & 2.5    \ 10$^{-14}$ \nl
834 & H$_2$ 0-0 S(0) & 28.218  & 2.3    \ 10$^{-14}$ \nl
668 & H$_2$ 0-0 S(1) & 17.035  & 2.8    \ 10$^{-14}$ \nl
834 & H$_2$ 0-0 S(1) & 17.035  & 2.9    \ 10$^{-14}$ \nl
668 & H$_2$ 0-0 S(3) & 9.66492 & $<$ 5.4\ 10$^{-15}$ \nl
668 & H$_2$ 0-0 S(5) & 6.90952 & $<$ 7.1\ 10$^{-15}$ \nl
\enddata
\end{deluxetable}
% -------------------------------------------------------------------

\newpage
\begin{figure}
\begin{center}
\leavevmode
\psfig{figure=ggtau_fig1.ps,height=18cm,angle=180}
\end{center}
\end{figure}
\end{document}